\documentclass[fleqn,10pt]{SelfArx} 

\definecolor{color1}{RGB}{0,0,90} 
\definecolor{color2}{RGB}{0,20,20} 

\usepackage{hyperref} 
\hypersetup{hidelinks,colorlinks,breaklinks=true,urlcolor=color2,citecolor=color1,linkcolor=color1,bookmarksopen=false,pdftitle={Title},pdfauthor={Author},urlcolor=blue}

\usepackage{graphicx}
\usepackage[square]{natbib}
\usepackage{amsmath}
\usepackage{multirow}
\usepackage{subfigure}
\usepackage{pdflscape}
\usepackage{graphicx}
\usepackage{natbib}
\usepackage{amsmath}
\usepackage{amsfonts}
\usepackage{multirow}
\usepackage{subfigure}
\usepackage{tikz}
\usetikzlibrary{calc}

\usepackage{mathtools,amssymb,lipsum}

\usepackage{cuted}
\usepackage{boondox-calo}

\newcommand{\bs}[1]{\boldsymbol{#1}}
\newcommand{\mc}[1]{\mathcal{#1}}
\newcommand{\nb}{\bs{\nabla}}

\newcommand{\n}[1]{\mathrm{#1}}

\JournalInfo{Published in Journal of Magnetism and Magnetic Materials, Vol. 535, 168057, 2021} 
\Archive{\href{https://doi.org/10.1016/j.jmmm.2021.168057}{DOI: 10.1016/j.jmmm.2021.168057}} 

\PaperTitle{MagTense: a micromagnetic framework using the analytical demagnetization tensor} 

\Authors{R. Bj\o{}rk, E. B. Poulsen, K. K. Nielsen, and A. R. Insinga} 
\affiliation{\textit{Department of Energy Conversion and Storage, Technical University of Denmark - DTU, Anker Engelunds Vej 1, DK-2800 Kgs. Lyngby, Denmark}} 
\affiliation{*\textbf{Corresponding author}: rabj@dtu.dk} 

\newcommand{\keywordname}{Keywords} 

\Abstract{We present the open source micromagnetic framework, MagTense, which utilizes a novel discretization approach of rectangular cuboid or tetrahedron geometry ``tiles'' to analytically calculate the demagnetization field. Each tile is assumed to be uniformly magnetized, and from this assumption only, the demagnetization field can be analytically calculated. Using this novel approach we calculate the solution to the $\mu$mag standard micromagnetic problems 2, 3 and 4 and find that the MagTense framework accurately predicts the solution to each of these. Finally, we show that simulation time can be significantly improved by performing the dense demagnetization tensor matrix multiplications using NVIDIA CUDA.}

\begin{document}

\flushbottom 

\maketitle 


\thispagestyle{empty} 

\section{Introduction}
At the microscopic scale, magnetic systems can be described by the formalism of micromagnetics. The purpose of this formalism is to determine the magnetization field within magnetic materials, and to calculate its time-evolution. Micromagnetic models are used for a plethora of magnetic calculations, including e.g. micromagnetics of rare-earth efficient permanent magnets \cite{Fischbacher_2018, gong2019a} and calculation of spin waves in nanometre-scale patterned magnetic elements \cite{Kim_2010}. The physics of micromagnetics can only in rare cases be solved analytically, and therefore numerical models are needed to simulate the physical behavior of the studied systems. Numerous micromagnetic models exist and some are also published as open source software \cite{Donahue_1999,Vansteenkiste_2014,Leliaert_2018}.

In a micromagnetic model, the governing equations must be discretized and represented by a finite number of degrees of freedom \cite{Kumar_2017}, and a number of distinct interactions takes place between the discretized moments in the systems. These interactions can be described by different magnetic fields, such as an exchange field and an anisotropy field. Of these it is the calculation of the demagnetization field, also known as the stray field, that is the most time consuming part \cite{Abert_2013}. There are in general three methods that have been applied to calculate the demagnetization field: finite difference based fast Fourier transform methods, tensor-grid methods and finite-element methods, although other methods such as fast multipole methods also exist \cite{Van_de_Wiele_2008,Palmesi_2017,Kumar_2017}. All methods have inherent advantages and shortcomings. The finite difference Fourier-based approaches are extremely fast but unfortunately require that the magnetic
moments lie on a regular grid \cite{Vansteenkiste_2011,Vansteenkiste_2014,Ferrero_2020}, although methods for non-uniform grids \cite{Livshitz_2009,Exl_2014} and multi-layer materials \cite{Lepadatu_2019} have also been proposed. The tensor-grid method \cite{Exl_2012} is still under investigation for stability and correctness \cite{Abert_2013}. The finite element method allows easy meshing of complex geometries but also requires that an environment outside the object of interest is modelled to ensure that the correct field is calculated.

So far the direct analytical calculation of the demagnetization field has only been explored in combination with a Fourier transform method \cite{Vansteenkiste_2011}. Here we present a novel micromagnetic model that uses an analytically exact approach to calculate the full demagnetization tensor for computations in real space. This ensures that the calculated magnetic field is correct, and at the same time allows for easy meshing of complex geometries. The disadvantage is an increase in the computational cost of the model.

\section{Physics of the model}

\subsection{Governing equations}\label{sec:GovEq}
In the following we will denote the magnetization field by $\bs{M}(\bs{x})$, where $\bs{x}$ denotes an arbitrary point. One of the fundamental assumptions in micromagnetism is that the norm of $\bs{M}$ is equal at all points to the saturation magnetization $M_s$. Therefore, it is customary to express the magnetization field as $\bs{M}(\bs{x}) = M_s \bs{m}(\bs{x})$, where $\bs{m}$ is the reduced magnetization that is normalized at any point. For heterogeneous materials the saturation magnetization can also be space-dependent, but it is always assumed to be a local known property of the magnetic material.

The equation of motion governing the dynamics of micromagnetic systems is the Landau-Lifshitz equation \cite{j2014a}:
\begin{equation}
\frac{d\bs{m}}{dt}  =-\gamma \;{\bs{m}}  \times \bs{H}_{\text{\scriptsize eff}} -\alpha \; {\bs{m}}  \times \left( {\bs{m}}  \times \bs{H}_{\text{\scriptsize eff}}\right) \label{eq:LLeq}\end{equation}
where $\bs{H}_{\text{\scriptsize eff}}$ denotes the effective field, a generalization of the concept of magnetic field that incorporates the effect of different interaction mechanisms. Indeed, the Landau-Lifshitz equation is a time-dependent partial differential equation within continuum physics that describes macroscopic classical effects as well as a semiclassical treatment of collective quantum effects.

The first and second terms on the right-hand side of Eq.~\ref{eq:LLeq} are called the precessional and damping term, respectively. In fact, at any point the reduced magnetization vector precesses around the effective field with a frequency determined by the gyromagnetic ratio $\gamma$ and rotates towards the effective field with a speed determined by the damping parameter $\alpha$. Due to the damping term, the cross product $\bs{m} \times \bs{H}_{\text{\scriptsize eff}}$ decreases over time, thus reducing the precession amplitude. At equilibrium, the magnetization vector is aligned with the effective field at any point. Therefore, the equilibrium state is described by the first Brown's equation:
\begin{equation}\bs{m} \times \bs{H}_{\text{\scriptsize eff}} = \bs{0}\end{equation}
As can be seen from Eq.~\ref{eq:LLeq}, when this condition is verified the magnetization is stationary: $\frac{d\bs{m}}{dt} =\bs{0}$.
Similarly, the second Brown's equation states that, at the border of the magnet
\begin{equation}\label{eq:Brown2}
    {\bf{m}}\times \left(A_{\text{\scriptsize exch}} \frac{\partial \bf{m}}{\partial \bf{n}}\right)=\bf{0},
\end{equation}
where $A_{\text{\scriptsize exch}}$ is the exchange interaction constant and $\bf{n}$ is the normal vector of the boundary. Since $\bf{m}$ and $\frac{\partial \bf{m}}{\partial \bf{n}}$ are orthogonal, Eq.~\ref{eq:Brown2} can only be fulfilled by $\frac{\partial \bf{m}}{\partial \bf{n}}=0$, which corresponds to Neumann boundary conditions.
It should also be stressed that the Landau-Lifshitz equation preserves the norm of $\bs{m}$ at any point.
\subsection{Free energy  and effective field}
The effective field is associated with the micromagnetic free energy of the system, denoted by $\mc{G}$. The free energy is expressed as the volume integral of a corresponding volumetric energy density $g$:
\begin{equation} \mc{G}[\bs{m}] = \int_{\Omega} dV \; g(\bs{x},\bs{m},\bs{\nb}m_x,\bs{\nb}m_y,\bs{\nb}m_z) \label{eq:EnergyDensity}\end{equation}
where $\Omega$ denotes the region occupied by the magnetic material. As can be seen from Eq.~\ref{eq:EnergyDensity}, the energy density may depend on the magnetization $\bs{m}$ and the spatial gradients of its $x$, $y$ and $z$ components. The effective field is proportional to the first variation of the free energy with respect to $\bs{m}$:
\begin{equation} \bs{H}_{\text{\scriptsize eff}} =-\frac{1}{\mu_0 M_s} \frac{\delta \mc{G}}{\delta \bs{m}} \label{eq:EffField}\end{equation}
where $\mu_0$ denotes the vacuum permeability. Therefore, the damping term in the Landau-Lifshitz equation has the effect of decreasing the free energy, and the minima of $\mc{G}$ with respect to $\bs{m}$ are equilibrium configurations, i.e. satisfying Brown's first equation.
\subsection{Interaction mechanisms}
As mentioned in Sec.~\ref{sec:GovEq}, the effective field is composed of several terms associated with different interaction mechanisms, each obtained from a corresponding term of the free energy by calculating the first variation.
\begin{itemize}
    \item \textbf{External field} - This term represents the classical macroscopic magnetostatic interaction of the magnetization vector with an external applied $\bs{H}_a$. The energy density is given by:
    \begin{equation}g_{\text{\scriptsize ext}} = -\mu_0 M_s (\bs{m} \cdot \bs{H}_a) \end{equation}
    The effective field associated with this interaction is simply coincident with the applied field itself:
    \begin{equation}\left(\bs{H}_{\text{\scriptsize eff}} \right)_{\text{\scriptsize ext}} = \bs{H}_a \end{equation}
    \item \textbf{Demagnetization field} - Similar to the interaction with the external field, the demagnetization term expresses the classical magnetostatic interaction of the magnetization with the field $\bs{H}_d$ generated in $\Omega$ by its own magnetization $\bs{M}(\bs{x})$:
    \begin{equation}g_{\text{\scriptsize demag}} = -\left(\frac{\mu_0 M_s}{2}\right)  (\bs{m} \cdot \bs{H}_d)\end{equation}
    The demagnetization field depends linearly on $\bs{m}$ according to the following expression:
    \begin{equation}\bs{H}_d (\bs{x}) = \int_{\Omega} dV'\, \underline{\underline{N}} (\bs{x},\bs{x}') \, \bs{m}(\bs{x}') \label{eq:DemagFieldCont}\end{equation}
    where $\underline{\underline{N}}$ denotes the demagnetization tensor. As for the previous term, the effective field associated with this interaction is coincident with the demagnetization field:
    \begin{equation}\left(\bs{H}_{\text{\scriptsize eff}} \right)_{\text{\scriptsize demag}} = \bs{H}_d\end{equation}
    The effect of this interaction is to reduce the field generated outside $\Omega$ by driving the magnetization vector at any point towards the local magnetic field, thus forming closed loops.
    \item \textbf{Exchange interaction} - This term is a semiclassical treatment of a collective quantum effect. Derived from the Heisenberg Hamiltonian, it expresses the tendency of the magnetization vector in a point to align to the magnetization vector of the surrounding points. In the semiclassical formalism this is achieved by minimizing the spatial derivatives of $\bs{m}$:
     \begin{equation}g_{\text{\scriptsize exch}} = A_{\text{\scriptsize exch}}\, \Big(\| \bs{\nb} m_x \|^2 + \| \bs{\nb} m_y \|^2 +\| \bs{\nb} m_z \|^2 \Big)\end{equation}
     where $A_{\text{\scriptsize exch}}$ denotes the exchange interaction constant.
     The effective field associated with the exchange interaction can be computed from the corresponding energy density using variational methods:
     \begin{equation}\left( \bs{H}_{\text{\scriptsize eff}}\right)_{\text{\scriptsize exch}} =\left( \frac{ 2 A_{\text{\scriptsize exch}} }{\mu_0M_s}\right)\left(\frac{\partial^2\bs{m}}{\partial x^2} + \frac{\partial^2\bs{m}}{\partial y^2} +\frac{\partial^2\bs{m}}{\partial z^2}\right).\end{equation}
     The effect of this interaction is thus analogous to that of a vector diffusion equation, i.e. driven by magnetization gradients across $\Omega$.
      As for $M_s$, $A_{\text{\scriptsize exch}}$ may be space-dependent for heterogeneous materials, in which case the effective field would be
      \begin{equation}\left( \bs{H}_{\text{\scriptsize eff}}\right)_{\text{\scriptsize exch}} =\left( \frac{ 2 }{\mu_0M_s}\right)\nabla\cdot \left(A_{\text{\scriptsize exch}}\nabla\bs{m}\right)\end{equation}
      and in case of an abrupt material change,
      \begin{equation}\label{eq:BrownEqAlt}
          A_{\text{\scriptsize exch}_1} \frac{\partial \bf{m}}{\partial {\bf{n}}_1}=A_{\text{\scriptsize exch}_2} \frac{\partial \bf{m}}{\partial {\bf{n}}_2},
      \end{equation}
      where $\bf{n}$ is the normal vector to the material interface. Eq.~\ref{eq:BrownEqAlt} is nothing but a generalization of the second Brown's equation, Eq.~\ref{eq:Brown2}. For simplicity we do not consider this dependence here, but it would be a relatively straightforward extension of the Neumann boundary conditions already present on the boundary of the magnet.
     \item \textbf{Crystal anisotropy} - This term describes the anistropic effects due to the local orientation of the crystal lattice. For the case of uni-axial anisotropy, the energy density is expressed as:
     \begin{equation} g_{\text{\scriptsize anis}} = -K (\bs{m}\cdot \hat{\bs{e}}_K)^2
     \end{equation}
     where $K$ denotes the anisotropy constant and $\hat{\bs{e}}_K$ is a unit vector oriented along the easy axis of the crystal. These two parameters can also be space-dependent. Typically, $\hat{\bs{e}}_K$ has a different direction in each crystal grain. The effective field associated to the uni-axial anisotropy term is given by:
      \begin{equation}\left( \bs{H}_{\text{\scriptsize eff}}\right)_{\text{\scriptsize anis}} = \frac{2 K}{\mu_0 M_s} (\bs{m}\cdot \hat{\bs{e}}_K) \hat{\bs{e}}_K. \end{equation}
      Therefore, the effect of this term is to rotate the magnetization vector towards the local direction of the easy axis, i.e. $\pm \hat{\bs{e}}_K$. When the anisotropy constant is negative, this term rotates $\bs{m}$ away from $\pm \hat{\bs{e}}_K$, towards the plane that is normal to $\hat{\bs{e}}_K$. In this case, $\hat{\bs{e}}_K$ and its normal plane are referred to as ``hard axis'' and ``easy plane'', respectively.
\end{itemize}
These four interactions are the most well studied mechanisms within the field of micromagnetics. However, it is possible to model other physical phenomena with a similar formalism, e.g. magnetostriction \cite{rotarescu2019a} or the Dzyaloshinskii-Moriya interaction \cite{perez2014a} (also called antisymmetric exchange interaction) or thermal fluctuations \cite{ragusa2009a}.

\section{Implementation: the MagTense framework}
Except for a few cases, most notably that of a uniformly magnetized particle described by the Stoner-Wohlfarth model, micromagnetics problems can only be approached by means of numerical computations. It is thus necessary to discretize the governing equations so that $\bs{m}(\bs{x})$ is represented by a finite number of degrees of freedom. As mentioned previously the most common discretization methods are finite elements methods and finite difference methods typically with a Fourier-based approach. Generally, the computation of the demagnetization field introduced in Eq.~\ref{eq:DemagFieldCont} is the most computationally intensive task in the calculation and evolution of $\bs{m}(\bs{x})$.

In this work, we present a novel open-source micromagnetism and magnetostatic framework MagTense, which uses neither the finite difference nor the finite element approach to calculating the demagnetization field, but instead relies on an analytically correct expression of the demagnetization tensor. The discretization approach used in MagTense subdivides the region $\Omega$ into $N$ ``tiles'', adjacent to each other. The magnetization field $\bs{m}(\bs{x})$ is thus represented by a vector of $3N$ components, three for each tile.

\begin{figure}[!t]
\centering
\includegraphics[width=0.48\textwidth]{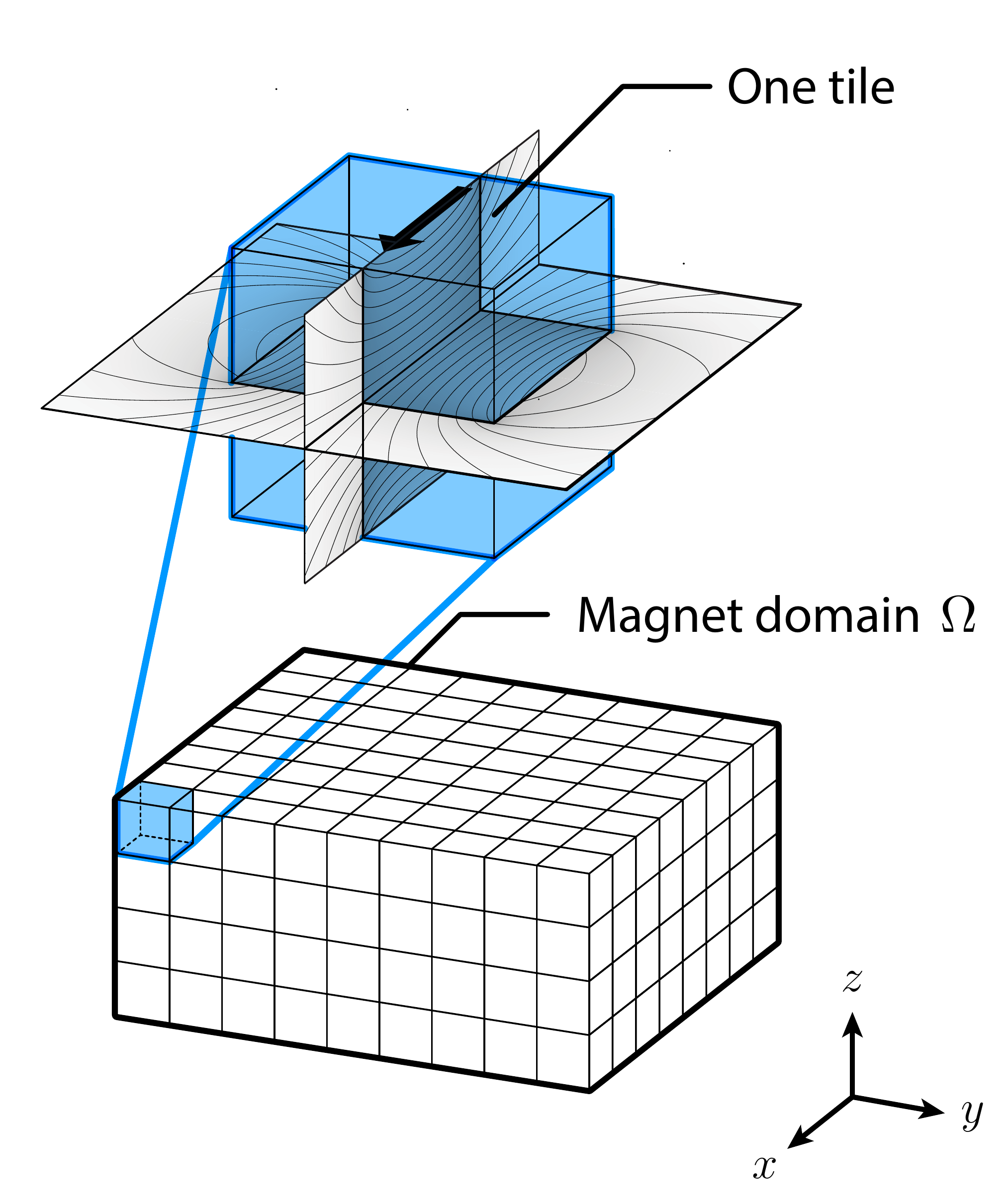}
\caption {An illustration of the tile concept used in MagTense. The magnetic field calculated by a single prism tile along an $xz$- and $xy$-plan is also illustrated.}\label{fig:Tiling}
\end{figure}

The demagnetization field generated by each of the tiles at the centers of all of the other tiles is computed analytically by assuming that the tile is uniformly magnetized. The analytical expression of the demagnetization tensor $\underline{\underline{N}}$ is known for many different shapes, such as the rectangular cuboid (rectangular prism) \cite{Smith_2010} and the tetrahedron \cite{nielsen2019a}, thus allowing great flexibility to build meshes that are suitable for the geometry being investigated. The total demagnetization field at any point is obtained by summing the individual contributions from all the tiles in the magnet domain $\Omega$. This concept is illustrated in Fig. \ref{fig:Tiling} for a magnet domain meshed as rectangular prisms.

Besides the calculation of the demagnetization field, the other components of the effective field must also be calculated, i.e. the exchange field, the external field and anisotropy field. For the case of a Cartesian mesh of rectangular cuboids the exchange interaction field can easily be computed by applying standard finite difference schemes. However, several methods are also available for the case of unstructured polyhedral meshes \cite{sozer2014a}. These will be discussed in a subsequent work, as determining the ideal differential operator for an unstructured mesh is not a trivial task. The implementation of the external field and anisotropy terms is straightforward, as these only depend on the local properties of the tile.

\subsection{The numerical implementation}
In the following, we will denote with the symbol $\bs{m}$ the $3N$-element vector representing the magnetization field over the discretized mesh. Since the exchange, anisotropy and demagnetization terms of the free energy are all quadratic with respect to $\bs{m}$, the corresponding terms of the effective field are linear. Therefore, they can all be computed by multiplying the $3N$-element vector by corresponding $3N \times 3N$ matrices \cite{insinga2020a}. The exchange and anisotropy matrices are sparse matrices, whereas the demagnetization matrix is dense, which is generally the reason why computing this term is the most intensive part of a micromagnetic calculation \cite{Abert_2013}. The external field energy is linear with respect to $\bs{m}$, implying that the corresponding effective field does not depend on $\bs{m}$. Denoting by $\underline{\underline{A}}$ the matrix associated with the quadratic terms, and by $\bs{b}$ the vector associated with the external field term, the $k^{\text{\scriptsize{th}}}$ component of the total effective field is computed as:
\begin{equation}
    \left( \bs{H}_{\text{\scriptsize eff}}\right)_{k} = \sum_{j=1}^{3N} A_{kj} m_{j}+b_k\quad\text{for}\;k=1,\dots,3N
\end{equation}

All the matrices need to be computed only once, at the beginning of the simulation. As the magnetization field evolves during the simulation according to Eq.~\ref{eq:LLeq}, the effective field is updated and the time-derivative of each component of $\bs{m}$ can be computed. The integration of the equation of motion is here performed with an explicit Runge-Kutta (4,5) formula.

The framework MagTense is available in both a version that performs the matrix calculations on the CPU, but also a version that uses CUDA for performing the dense matrix-multiplication needed to calculate the demagnetization field, for increasing the calculation speed \cite{Lopez_Diaz_2012,Leliaert_2019}. All examples presented here as well as all model software is available from www.magtense.org \cite{MagTense}. The MagTense micromagnetic model is available both in the FORTRAN programming language and Matlab computing environment. The FORTRAN version can be directly interfaced with Matlab through a MEX-interface. A Python interface is available for the magnetostatic part of MagTense at present.

\subsection{Computational resources}
The challenge in using the full demagnetization tensor for micromagnetic simulations is that the demagnetization tensor scales as $N^2$ where $N$ is the number of tiles. This is in contract to e.g. the FFT method, where the storage requirement is only of size $N$ and the computational complexity is $N\mathrm{log}(N)$. For a micromagnetic problem with periodic boundary conditions the use of the direct demagnetization tensor has been shown to being computationally significantly slower than the evaluation of the tensor using FFT, as expected from this scaling \cite{Kruger_2013}. The analytical formulation of the demagnetization tensor can also lead to numerical errors in the evaluation of the tensor \cite{Kruger_2013}, with the error at large distances being of the same magnitude as the elements of the tensor itself \cite{Lebecki_2008}, but this is mostly a problem for sample with periodic boundaries, i.e. samples of infinite size, as comparisons for finite-sized samples have shown that the direct calculation of the demagnetization field ensured the lowest error of the various demagnetization field calculations methods \cite{Abert_2013}.

To reduce the problem of the $N^2$ scaling of the demagnetization tensor, MagTense utilizes the fact that the tensor is symmetric \cite{Moskowitz_1966}, reducing the scaling to $(N(N+1)/2)^2$ which is almost $N^2/2$. Furthermore, the demagnetization tensor is stored as a float (4 byte) and not a double (8 byte) also reducing the required memory.

However, it is clear that for large-scale micromagnetic simulation the number of elements in the demagnetization tensor will likely have to be reduced, either though a smart meshing of the structure to be simulated or through removing selected entries in the demagnetization tensor. The latter approach will be discussed subsequently in the next section.

\section{Verification of the MagTense Framework}
To verify the MagTense micromagnetic framework and the use of the full demagnetization tensor to compute the effective field, we consider three of the standard problems in micromagnetics as described in the $\mu$mag standard problems \cite{NIST_2020}. All the models described below are directly available as part of the MagTense framework \cite{MagTense}. For each standard problem the accuracy of the model is assessed, and in a single case the computation time of the model is also evaluated.

\begin{figure*}[!t]
\subfigure[]{\includegraphics[width=0.49\textwidth]{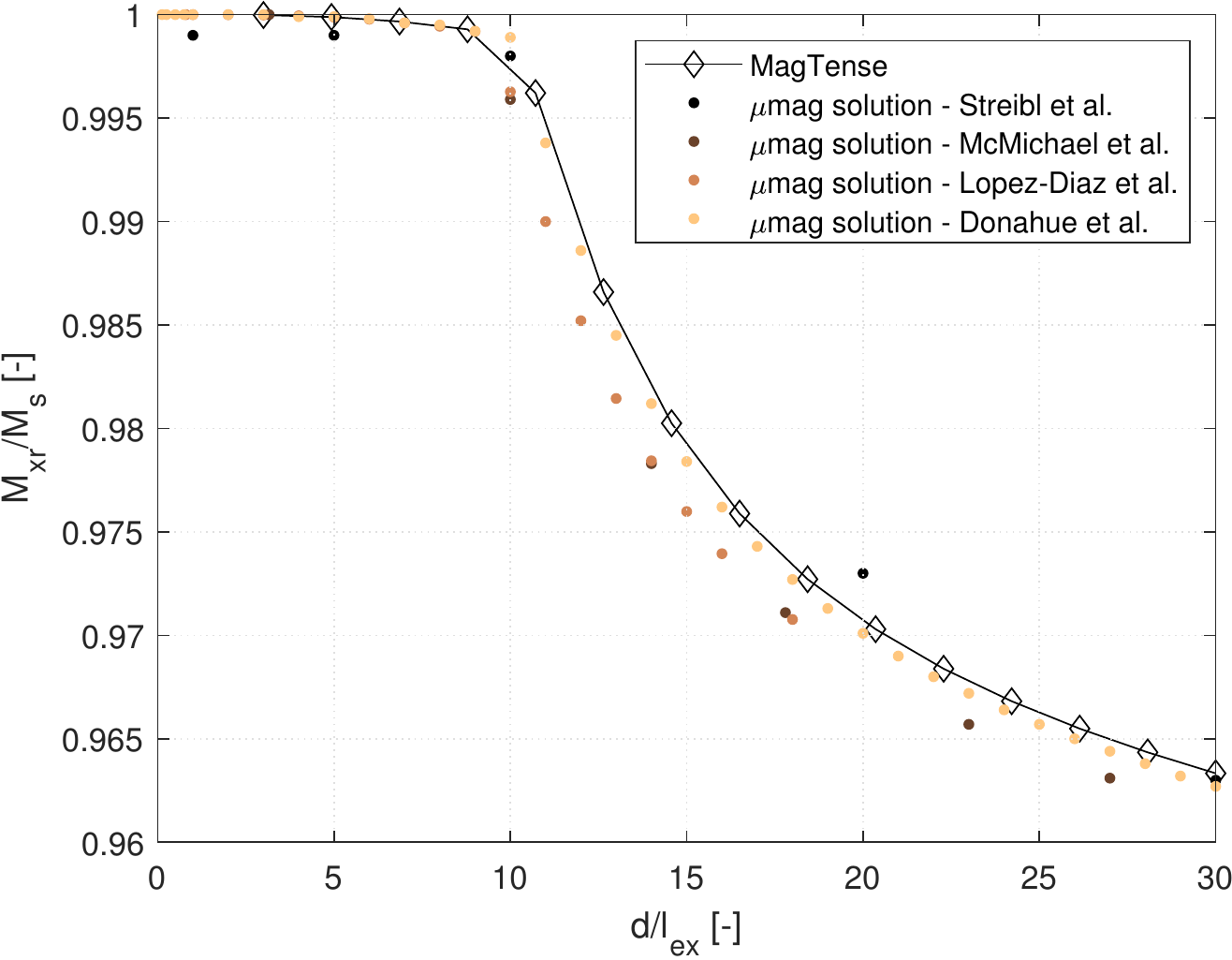}}
\subfigure[]{\includegraphics[width=0.49\textwidth]{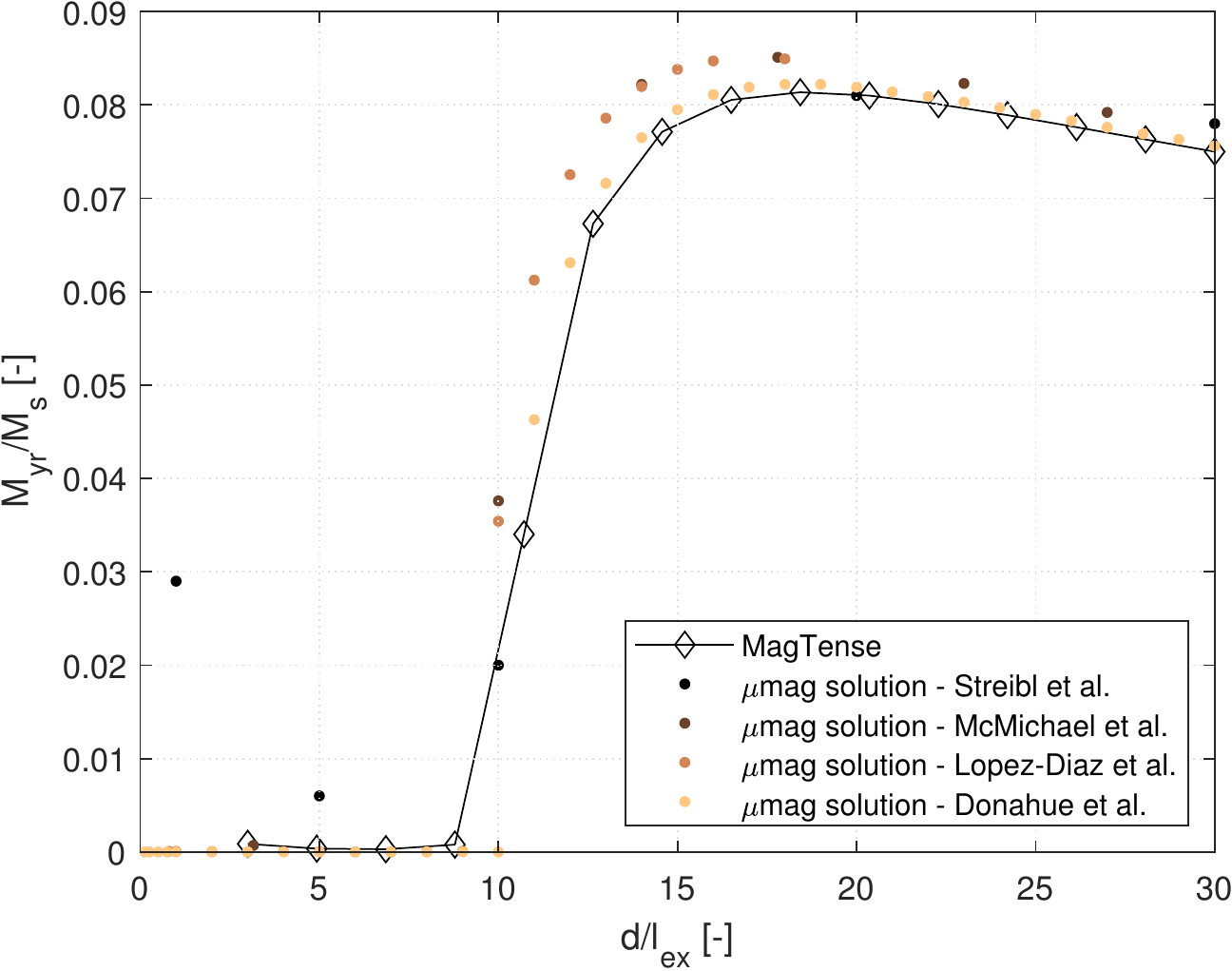}}
\subfigure[]{\includegraphics[width=0.49\textwidth]{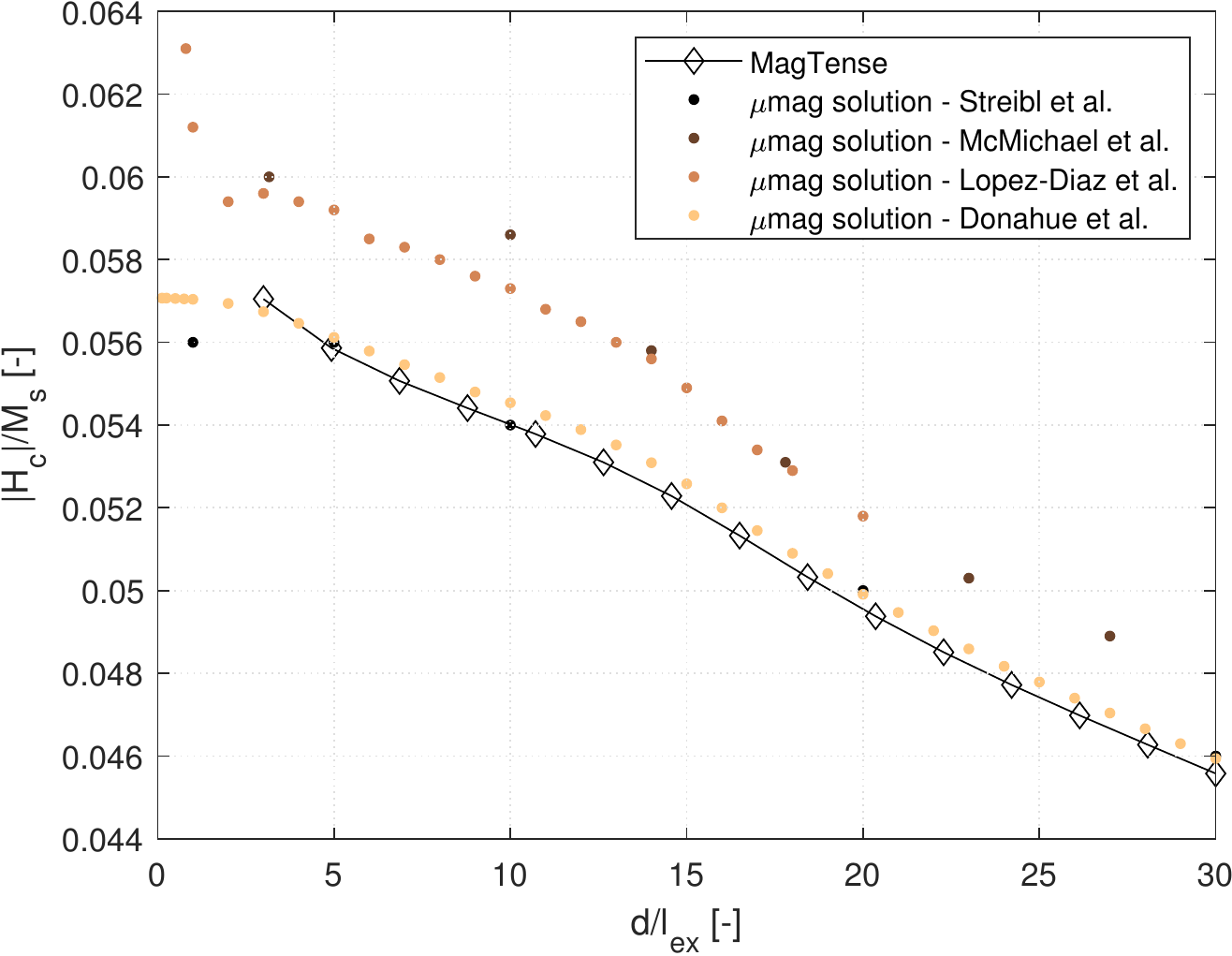}}
\subfigure[]{\includegraphics[width=0.49\textwidth]{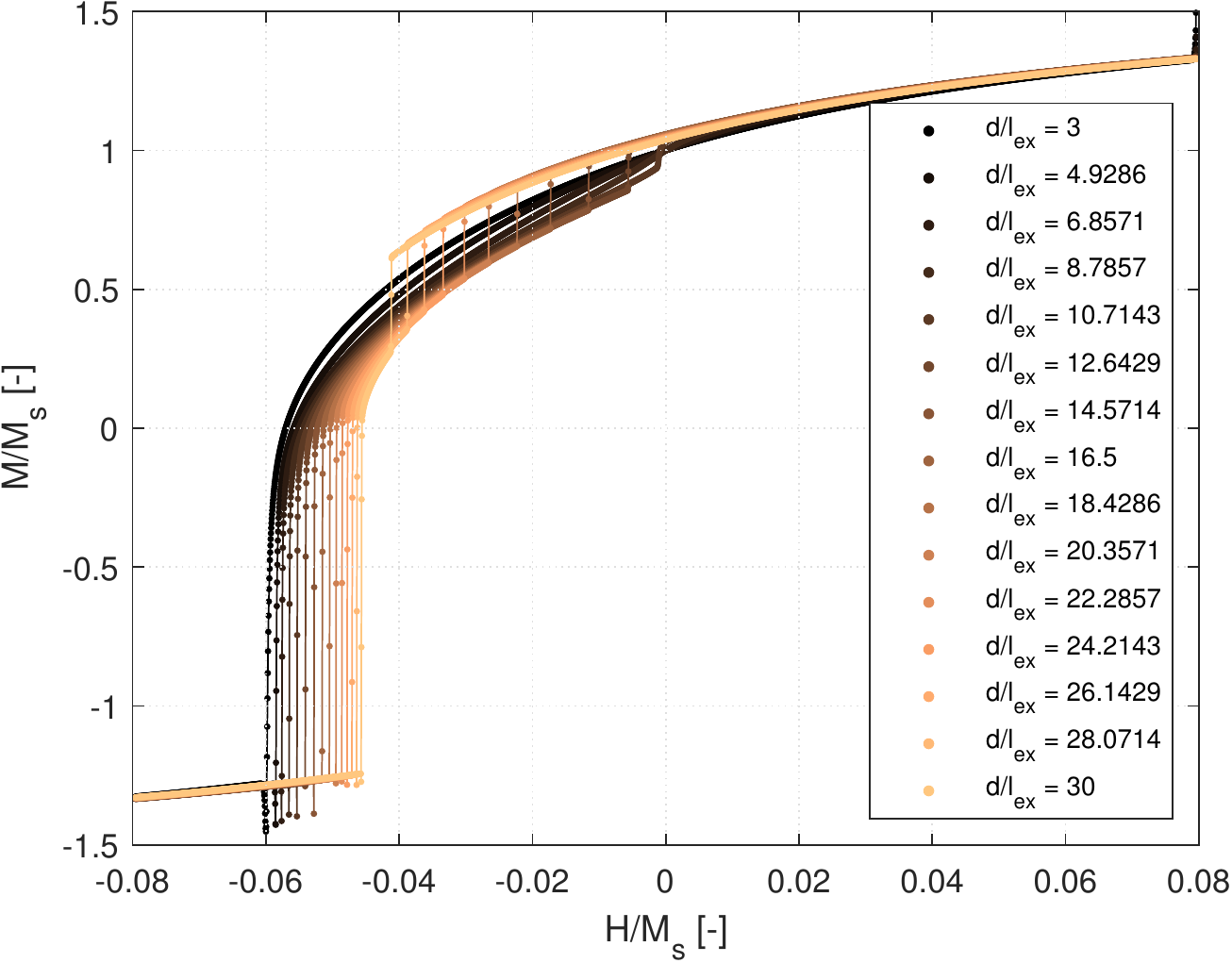}}
\caption{a) The remanent magnetization along the long axis as a function of $d/l_\mathrm{ex}$ and b) the remanent magnetization along the short axis as a function of $d/l_\mathrm{ex}$ and c) the coercivity for fields applied along the (111) direction as a function of $d/l_\mathrm{ex}$ and d) the computed hysteresis curves.}\label{Fig_problem_2}
\end{figure*}

\subsection{$\mu$mag standard problem 2}
In standard problem 2, hysteresis curves must be calculated for a thin film of length $L$, width $d = L/5$ and thickness $t = d/10$. Crystalline anisotropy is neglected in this problem. An external field is applied in the [1,1,1] direction, and the magnitude of the field is slowly decreased to calculate the hysteresis curve \cite{Streibl_1999}.
The hysteresis loop will depend on the scaled geometry of the problem, as the scale of the thin film dimensions change relative to the length scale of the exchange interaction. The results are specified in terms of the ratio between the thin film width $d$ and the exchange length $l_\mathrm{ex} = \sqrt{A_\mathrm{exch}/K_\mathrm{m}}$, where $A_\mathrm{exch}$ is the exchange stiffness constant and $K_\mathrm{m}$ is the magnetostatic energy density, $K_\mathrm{m} = M_\mathrm{s}^2/2\mu_0$.

In MagTense we utilize a Cartesian mesh with the number of rectangular cuboid tiles equal to $100\times{}20\times{}3$ in the $x$-, $y$- and $z$-direction respectively. We calculate a hysteresis loop with an applied field starting at 0.08 $H/M_\mathrm{s}$ and decreasing the field in 4000 steps to a value of -0.08 $H/M_\mathrm{s}$. The micromagnetic damping parameter, $\alpha$, and the precession parameter, $\gamma$, are not specified in the problem statement. Here we take the value of $\alpha=4000$, which is close to the value for standard problem 4, while we take $\gamma=0$, as we are only interested in the final steady state configuration. At each value of the field, we evolve the magnetization 40 ns to find the new equilibrium configuration of the magnetization, before the external field is changed again. These 40 ns are sufficient to reach the equilibrium configuration for almost all cases considered. The aim is to calculate the coercivity, which is the value of the field $H_\mathrm{c}/M_\mathrm{s}$ at which the projection of the magnetization along the field is zero, and the remanence, which is the value of $\vec{M}/M_\mathrm{s}$ at $H_\mathrm{a} = 0$.

It should be note that taking $\gamma=0$ is a valid approximation, even if the Landau-Lifshitz-Gilbert (LLG) equation, i.e. not the Landau-Lifshitz equation presented in Eq. \ref{eq:LLeq}, is considered. This is because in the limit of large damping coefficients $\alpha$ the gyromagnetic factor in the LLG scales as $\gamma/\alpha^2$ , whereas the damping factor scales as $\gamma/\alpha$. As the damping coefficient tends to infinity, the gyromagnetic factor then tends to 0 much faster than the damping does. Thus the apporach of using $\gamma=0$ is valid for this standard problem.

Shown in Fig. \ref{Fig_problem_2} is the thin film coercivity and remanence as a function of $d/l_\mathrm{ex}$ both for the MagTense model as well as results published in the $\mu$mag problem solutions \cite{NIST_2020}. As can be seen from the figure, the MagTense model is in agreement with the majority of the published $\mu$mag results for both the coercivity and the remanence values. In MagTense both of these values converge the slowest at low values of $d/l_\n{ex}$. This can be seen in Fig. \ref{Fig_problem_2}d, where at the start of the coercivity curve, at 0.08 $H/M_\mathrm{s}$, the simulation for the lowest $d/l_\mathrm{ex}$ requires a number of steps in the external field before the initial magnetization configuration has reached the correct equilibrium state. This is in line with previous results, which have demonstrated that in the small particle limit convergence can be difficult to obtain \cite{Donahue_2000}.

Here, the difficulty of obtaining converging at the small particle limit is most likely caused by the chosen ODE45 solver, which is not well suited to stiff problems. The demagnetization tensor approach used is correct also for length scales smaller than the exchange length, so it is not an imprecise calculation of the demagnetization field that is causing the slow convergence. An additional factor in the slow convergence at small length scales could be the chosen discretization of the finite difference exchange operator. This is chosen to be a 3-element stencil, which is correct to second order, but when the exchange coupling is strong a higher order operator may be needed. It has also previously been shown that the spatial discretization with an analytical demagnetization calculations can result in the Landau-Lifshitz equation becoming stiff \cite{Shepherd_2014}.

\subsection{$\mu$mag standard problem 3}
In standard problem 3, the goal is to calculate the minimum energy state in the single domain limit of a cubic magnetic particle of side length $L$. The standard problem consists on calculating the energy of two different states, known as the flower state and the vortex state respectively, to determine the value of $L$ at which there is a change in the minimum energy state of the cube from one of the states to the other.

The edge length of the considered cube, $L$, is similarly to standard problem 2 expressed in terms of the exchange length scale, $l_\n{ex} = \sqrt{A_\n{exch}/K_\n{m}}$, where $K_\n{m}$ is the magnetostatic energy density given by $K_\n{m} = M_\mathrm{s}^2/2\mu_0$. A uniaxial anisotropy with magnitude $K_\n{u}=0.1K_\n{m}$ along a principal axis of the cube is assumed.

\begin{figure}[!t]
\includegraphics[width=1\columnwidth]{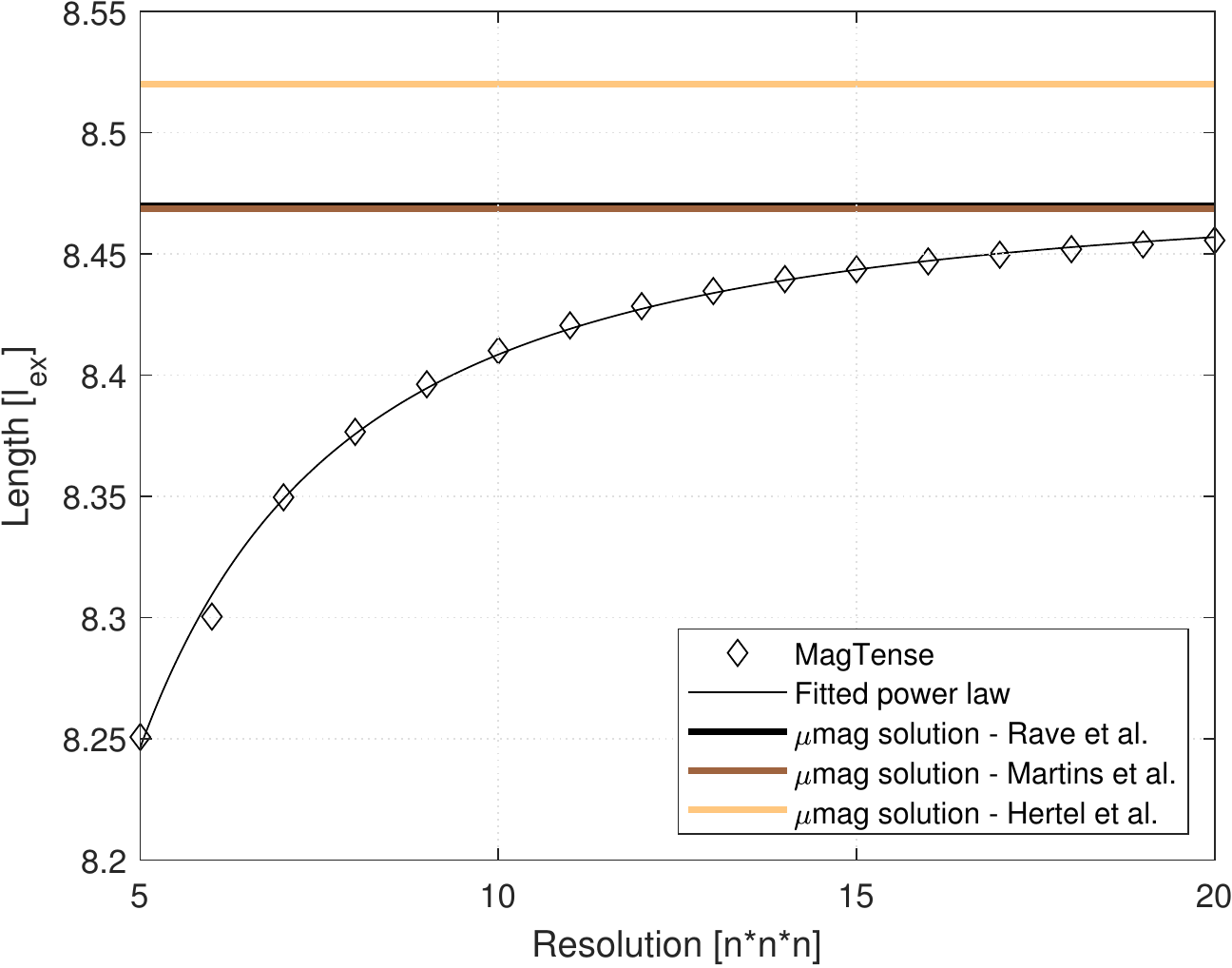}
\caption{The side length at which the energy crossover between the flower and vortex state happens, as a function of resolution, i.e. the number of tiles, $n$, in each direction. The tabulated $\mu$mag solutions, which are extrapolated to an infinitely fine grid, are shown as well. Two of these, the Rave et al. and Martins et al. overlap.}\label{Fig_problem_3_L}
\end{figure}

Shown in Fig. \ref{Fig_problem_3_L} is the side length in units of $l_\n{ex}$ at which the minimum energy state switches between the flower and the vortex state as a function of resolution, i.e. the number of cubic magnetic tiles in the model. As can be seen from the figure, the crossover points approach two of the three tabulated $\mu$mag solutions as the resolution is increased. Extrapolating the energy crossover to an infinitely fine resolution by fitting a power law of $L_\n{cross} = an^b+c$ to the data gives $L_\n{\infty}= 8.477 \pm 0.007$, which is in line with two of the published $\mu$mag results.

\begin{figure}[!t]
\includegraphics[width=1\columnwidth]{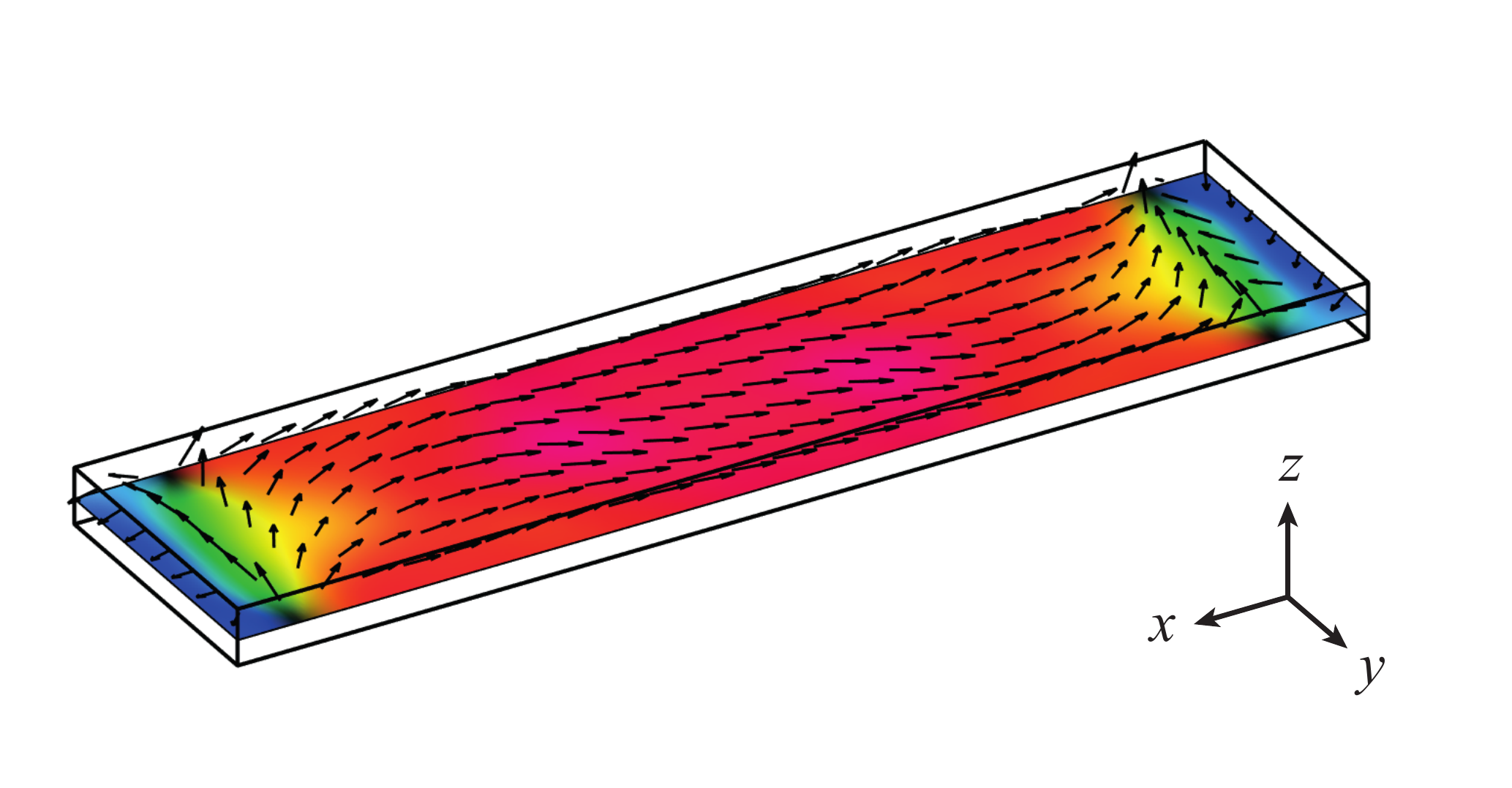}
\caption{State of the system at an intermediate step of the relaxation process for applied field 2 in standard problem 4. The magnetization direction in each point is indicated by the black arrows and by the color. In the figure, the thickness of the film in the $z$ direction has been exaggerated for clarity.  }\label{Fig_problem_4_OneStep}
\end{figure}

\subsection{$\mu$mag standard problem 4}
In standard problem 4 the dynamic evolution of a film with a thickness of 3 nm, a length of 500 nm and a width of 125 nm is considered. We assume that the problem can be discretized in a two-dimensional grid, i.e. that the thickness is only resolved by a single tile.

\begin{figure*}[!t]
\subfigure[]{\includegraphics[width=0.49\textwidth]{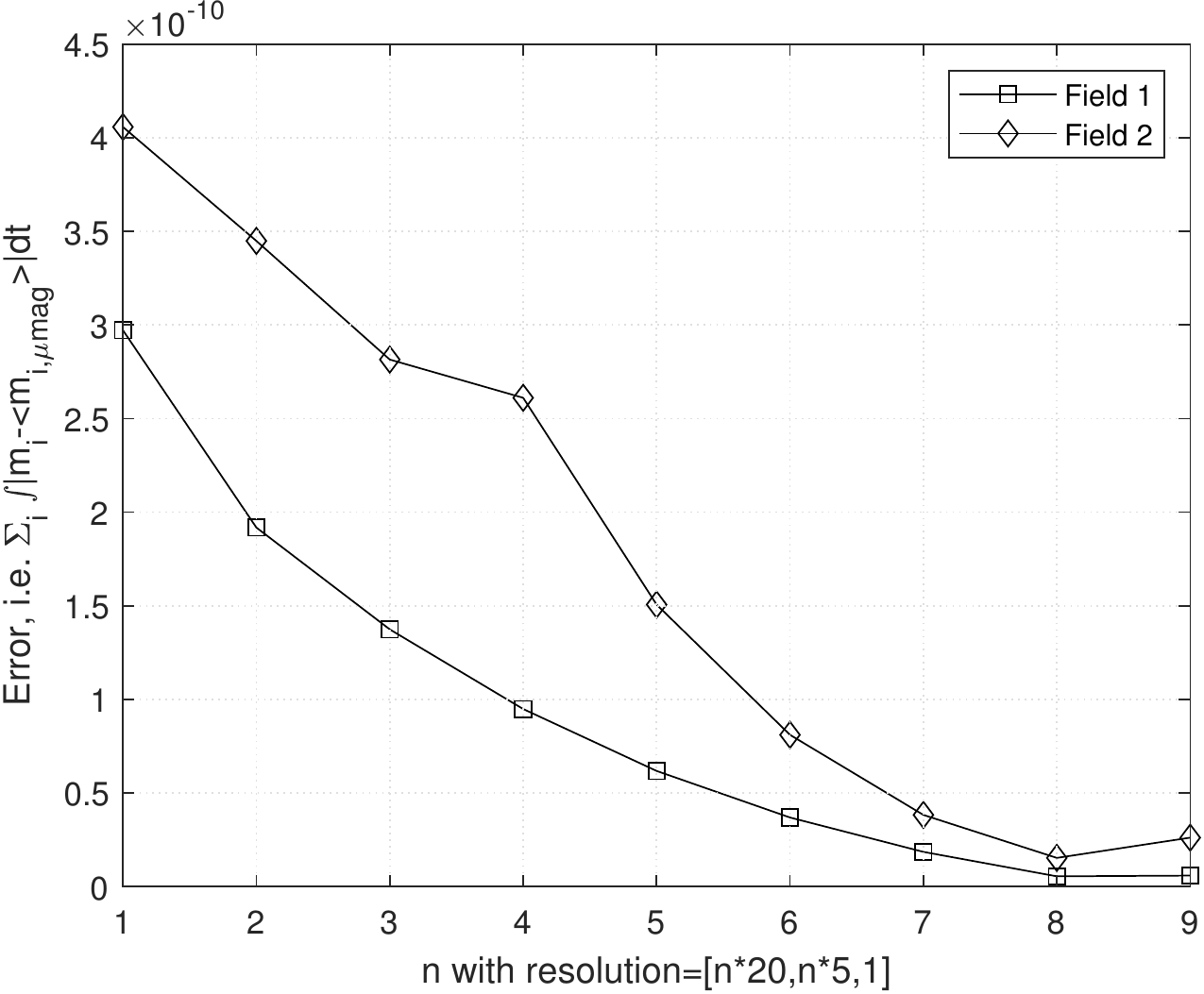}}
\subfigure[]{\includegraphics[width=0.49\textwidth]{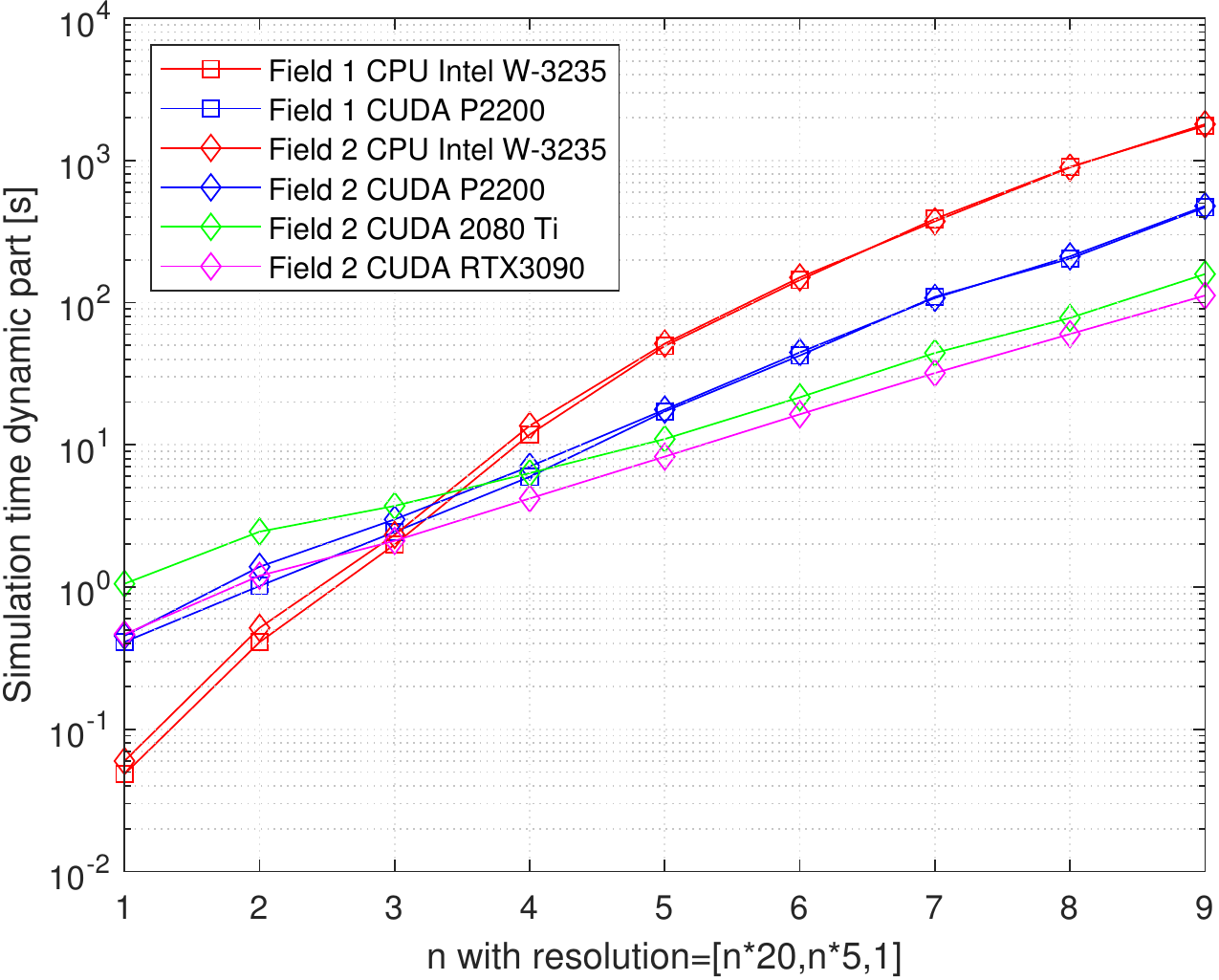}}
\caption{a) The error for standard problem 4 as defined in the text as a function of resolution and b) the computation time as a function of resolution on different hardware. }\label{Fig_problem_4_time_error}
\end{figure*}

First an initial equilibrium $s$-state must be calculated, which is obtained by first applying a saturating magnetic field along the [1,1,1] direction and then slowly removing this field and letting the magnetization equilibrate. Following this one of two different magnetic fields is applied instantaneously at $t=0$ to the initial $s$-state and the time evolution of the magnetization as the system moves towards equilibrium in the new fields is simulated. The two fields considered are field 1: $\mu{}_{0}H= [-24.6, 4.3,  0.0]$ mT and field 2: $\mu{}_{0}H= [-35.5, -6.3, 0.0]$ mT.

The parameters relevant for the problem are an exchange constant of $A_\n{exch} = 1.3\cdot10^{-11}$ J/m, a saturation magnetization of $M_\n{s} = 8.0\cdot10^5$ A/m, a damping parameter of $\alpha = 4.42\cdot10^3$ and a precession parameter of $\gamma = 2.21\cdot10^5$.

Using MagTense we compute the initial $s$-state and the following dynamical evolution for various resolutions of the uniform grid. The resolution is varied as $[20n,5n,1]$ with values of $n=1$ to $9$. Fig. \ref{Fig_problem_4_OneStep} shows an example of magnetization distribution over the cross-section of the film. This configuration corresponds to applied field 2 after $t=0.5$~ns from the beginning of the relaxation process.

\begin{figure}[!b]
\includegraphics[width=1\columnwidth]{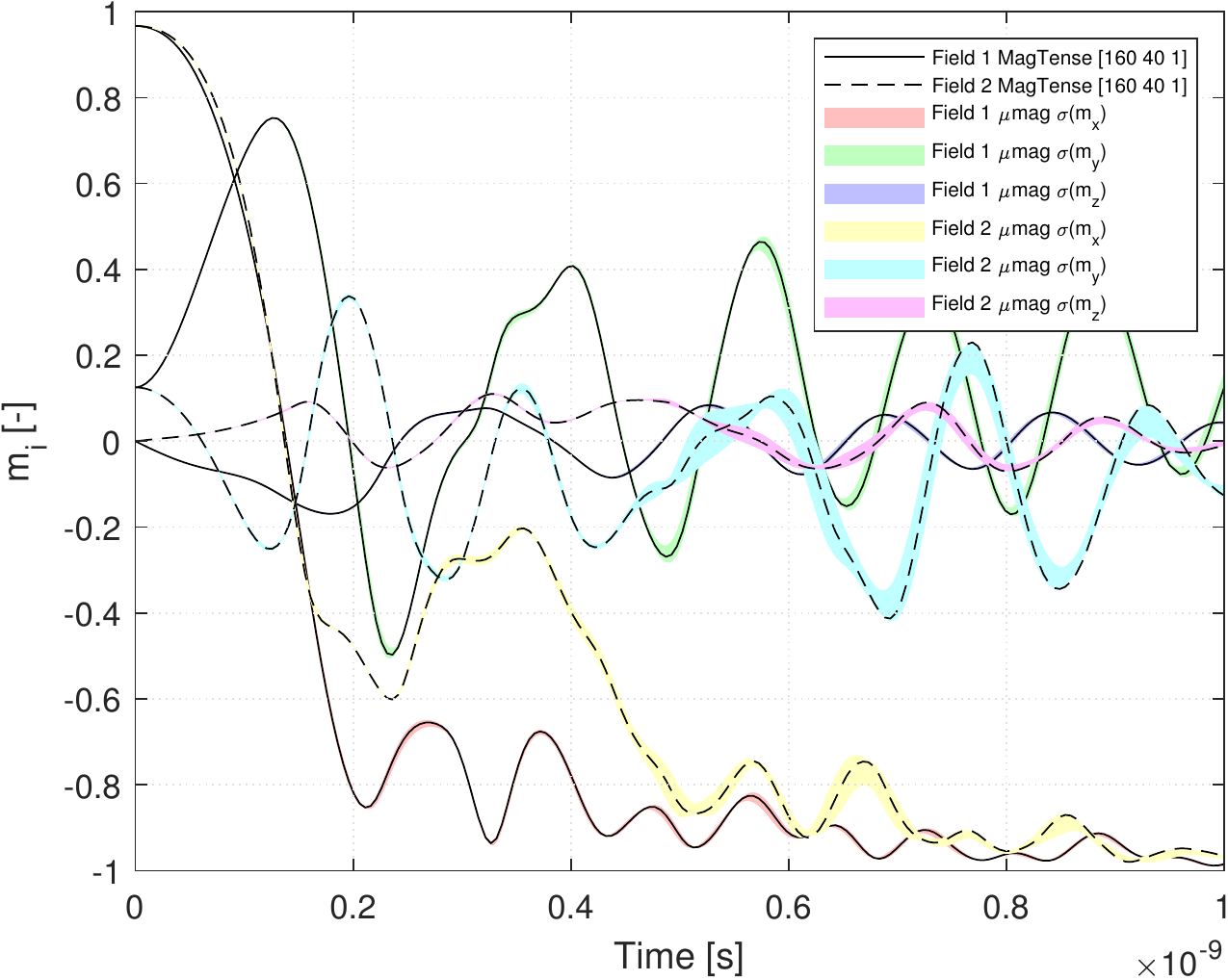}
\caption{The three components of the magnetization as a function of time for field 1 and 2 in standard problem 4 for a resolution of [160 40 1] along with the standard deviation of the published $\mu$mag solutions.}\label{Fig_problem_4_specific}
\end{figure}

Computing the mean of previously published models as tabulated in the $\mu$mag results, the error for the MagTense model can be defined as \begin{equation} err=\sum_\n{i}\int_{0}^{100\mathrm{ns}}\lvert m_\n{i}-\langle m_\n{i,\mu\mathrm{mag}}\rangle\rvert\mathrm{d}t, \end{equation} summing the different magnetization components and integrating over the first 100 ns of the time evolution. In Fig. \ref{Fig_problem_4_time_error}a the error is shown as a function of resolution.
As can be seen from the figure, the error decreases as the mesh resolution is increased, as expected. At a resolution of [160,40,1], i.e. $n=8$, the model has clearly converged, resulting in a minimum error between the model and the $\mu$mag tabulated data. In Fig. \ref{Fig_problem_4_time_error}b the simulation time for the dynamic part of standard problem 4, i.e. without the calculation of the initial s-state, is shown as a function of resolution for different hardware configuration. The cases considered are a simulation where all computations are done on an Intel W-3235 CPU, as well as calculations where the demagnetization field matrix multiplications are done using CUDA on a P2200, 2080 TI and a RTX 3090 Nvidia GPU respectively. As can be seen using CUDA for the calculations results in a significant improvement in computation time, with a similar magnitude as previously seen using CUDA micromagnetic frameworks \cite{Lopez_Diaz_2012,Leliaert_2019}.

\begin{figure*}[!t]
\subfigure[]{\includegraphics[width=0.49\textwidth]{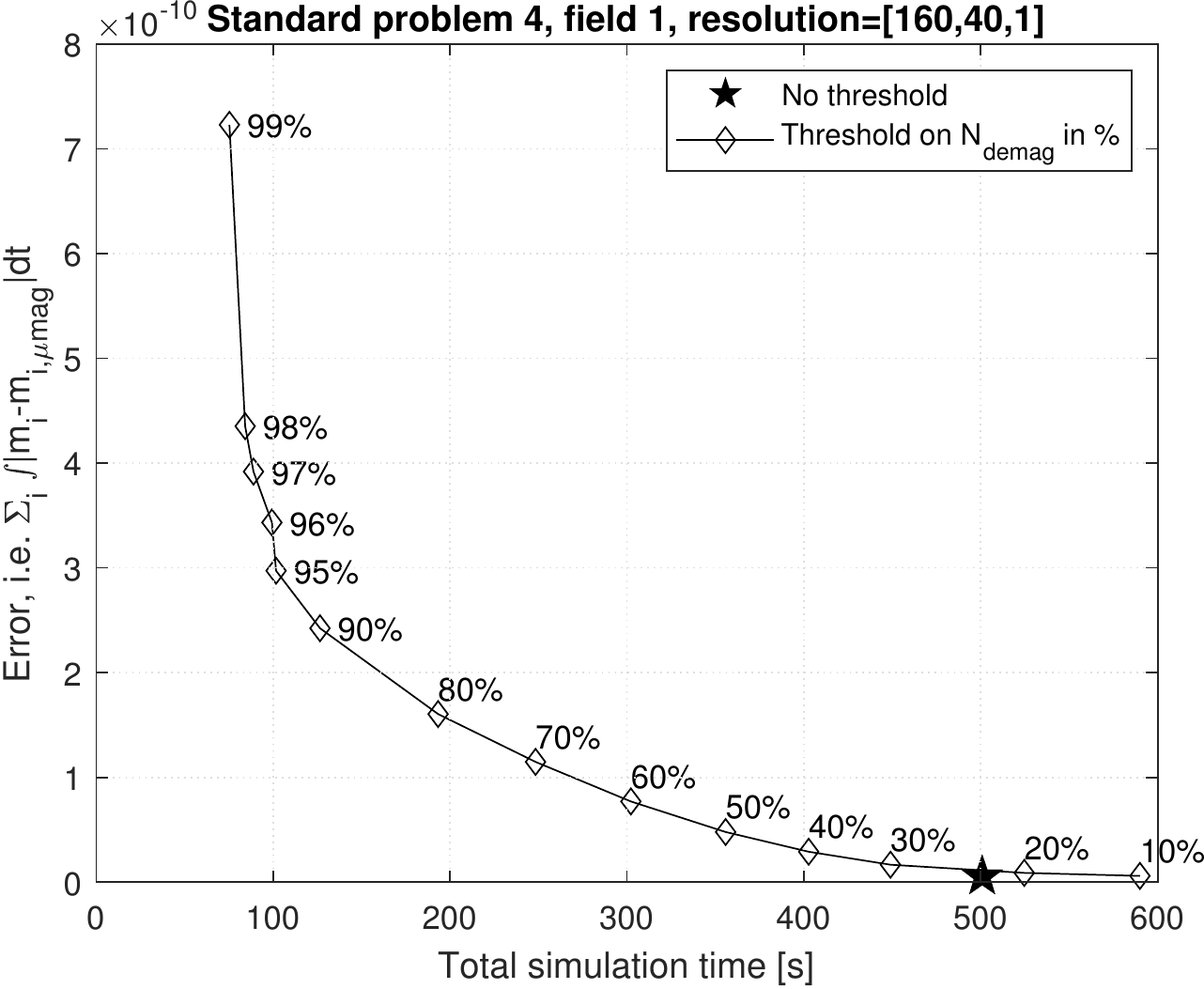}}
\subfigure[]{\includegraphics[width=0.49\textwidth]{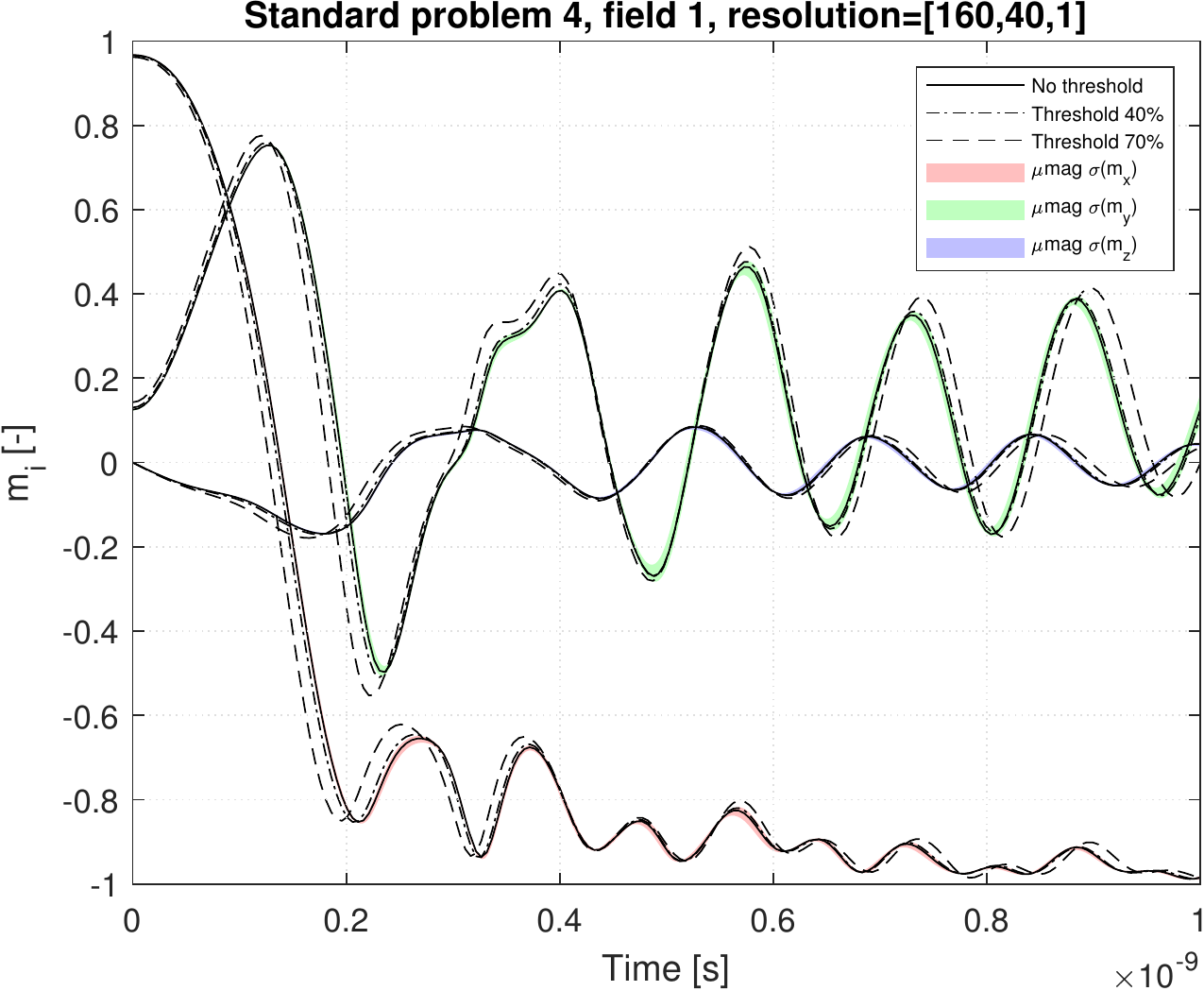}}
\caption{a) The error as a function of computation time for various threshold percentages, i.e. which percentage of the smallest values in the demagnetization tensor that is removed and b) the three components of the magnetization as a function of time for field 1 in standard problem 4 for various threshold fractions along with the standard deviation of the published $\mu$mag solutions. }\label{Fig_problem_4_threshold}
\end{figure*}

The dynamic evolution of the system is shown Fig. \ref{Fig_problem_4_specific} for a resolution of [160,40,1] along with the standard deviation of the published $\mu$mag solutions. As can be seen MagTense predicts an evolution that clearly is in line with the existing published $\mu$mag data.

\subsubsection{Influence of thresholding the demagnetization tensor}
It is of interest to explore whether or not the demagnetization tensor can be reduced in complexity from a fully dense matrix to a sparse one in order to improve the speed and memory use of the simulation. We have therefore implemented a feature in the MagTense framework such that the  demagnetization tensor can be reduced in size by removing all elements of the tensor smaller than a given threshold value, i.e. all values below the threshold are truncated. This also allows the demagnetization tensor, which normally is a dense matrix, to be allocated as a sparse matrix, thus saving memory, which is especially important when doing CUDA calculations. The effect of adjusting the threshold value for the computed solution to standard problem 4 is investigated in the following.

We consider a resolution of [160,40,1]. A threshold is chosen and the values which in absolute value is smaller than the threshold is set to zero. For example for a threshold of $20\%$ the absolute smallest fifth of the elements larger than zero in the tensor is truncated to zero. In Fig. \ref{Fig_problem_4_threshold}a the reduction in simulation time versus the error is shown for various values of the threshold, with the calculations performed on a Intel W-3235 CPU. As can be seen a reduction in computation time is possible without sacrificing accuracy. The computed dynamic solution is shown in Fig. \ref{Fig_problem_4_threshold}b for a threshold value of 40\% and 70\%, respectively, to illustrate that the solution, especially for the 40\% case, is still accurate compared to the tabulated $\mu$mag solutions. Note that for small values of the threshold, the calculation time increases because when a threshold is applied the demagnetization tensor is cast as a sparse matrix. When the number of elements in the tensor is still close to the fully dense matrix, there is a significant overhead in the sparse matrix multiplication of the demagnetization tensor with the magnetization, leading to the increase in simulation time.

It should be noted that thresholding the demagnetization tensor is the simplest and crudest method of reducing the number of entries in the tensor and is a technique that should be used with caution.
As the magnitude of the field of a single tile decreases as $\sim 1/r^3$ away from the tile, this truncation strategy will in essence result in that tiles only "see" other tiles in a more or less spherical region around themselves. This can be problematic as the number of tiles within such a sphere with radius $r$ grows as $r^3$, thus in effect cancelling out the $1/r^3$ decreasing influence of the individual tiles, making the demagnetization tensor important across the whole sample. A better but more complicated strategy for removing elements from the demagnetization tensor is to employ a multi-grid approach that scales the grid with the distance from a respective tile or use an adaptive mesh refinement technique \cite{Ramstock_1996,Hirano_1999,Garcia_2006,Sun_2006}. Another approach is to uses different approximations for the demagnetization tensor depending on the distance to the respective tile \cite{Lebecki_2008,Kruger_2013}.

Another strategy could be to calculate the full demagnetization tensor and subsequently perform an FFT transformation of it and use this matrix in calculations, transforming the magnetization to Fourier-space and back in the matrix multiplication with the demagnetization tensor. For the currently used regular grids this will make the method identical to the existing FFT frameworks, but for irregular grids this may be a computationally efficient strategy.

\section{Conclusion}
We have presented the MagTense micromagnetic framework, which utilizes a discretization approach of tiles of either rectangular cuboid or tetrahedron geometry to analytically calculate the demagnetization field. The only assumption in this approach is that a tile is uniformly magnetized.

Using this novel approach, which differs from existing finite difference and finite element schemes, we calculated the solution to a range of the $\mu$mag standard problems in micromagnetism and showed that the MagTense framework accurately calculates the solution to each of these.

Finally, we explored the simulation time in MagTense and showed that this can be significantly improved by performing the dense demagnetization tensor matrix multiplications using CUDA, and also that the demagnetization tensor can be reduced in complexity to increase the calculation speed without significantly sacrificing the precision of the solution, albeit with possible caveats depending on the shape of the sample.

\section*{Acknowledgment}
This work was supported by the Poul Due Jensen Foundation project on Browns paradox in permanent magnets, Project 2018-016.


\begin{thebibliography}{38}
\providecommand{\natexlab}[1]{#1}
\providecommand{\url}[1]{\texttt{#1}}
\expandafter\ifx\csname urlstyle\endcsname\relax
  \providecommand{\doi}[1]{doi: #1}\else
  \providecommand{\doi}{doi: \begingroup \urlstyle{rm}\Url}\fi

\bibitem[Fischbacher et~al.(2018)Fischbacher, Kovacs, Gusenbauer, Oezelt, Exl,
  Bance, and Schrefl]{Fischbacher_2018}
Johann Fischbacher, Alexander Kovacs, Markus Gusenbauer, Harald Oezelt, Lukas
  Exl, Simon Bance, and Thomas Schrefl.
\newblock Micromagnetics of rare-earth efficient permanent magnets.
\newblock \emph{Journal of Physics D: Applied Physics}, 51\penalty0
  (19):\penalty0 193002, 2018.

\bibitem[Gong et~al.(2019)Gong, Yi, and Xu]{gong2019a}
Qihua Gong, Min Yi, and Bai~Xiang Xu.
\newblock Multiscale simulations toward calculating coercivity of nd-fe-b
  permanent magnets at high temperatures.
\newblock \emph{Physical Review Materials}, 3\penalty0 (8):\penalty0 084406,
  2019.
\newblock ISSN 24759953.
\newblock \doi{10.1103/PhysRevMaterials.3.084406}.

\bibitem[Kim(2010)]{Kim_2010}
Sang-Koog Kim.
\newblock Micromagnetic computer simulations of spin waves in nanometre-scale
  patterned magnetic elements.
\newblock \emph{Journal of Physics D: Applied Physics}, 43\penalty0
  (26):\penalty0 264004, 2010.

\bibitem[Donahue and Porter(1999)]{Donahue_1999}
M.J. Donahue and D.G. Porter.
\newblock Oommf user's guide, version 1.0, interagency report nistir 6376.
\newblock Technical report, National Institute of Standards and Technology,
  Gaithersburg, MD, 1999.

\bibitem[Vansteenkiste et~al.(2014)Vansteenkiste, Leliaert, Dvornik, Helsen,
  Garcia-Sanchez, and Van~Waeyenberge]{Vansteenkiste_2014}
Arne Vansteenkiste, Jonathan Leliaert, Mykola Dvornik, Mathias Helsen, Felipe
  Garcia-Sanchez, and Bartel Van~Waeyenberge.
\newblock The design and verification of mumax3.
\newblock \emph{AIP advances}, 4\penalty0 (10):\penalty0 107133, 2014.

\bibitem[Leliaert et~al.(2018)Leliaert, Dvornik, Mulkers, De~Clercq,
  Milo{\v{s}}evi{\'c}, and Van~Waeyenberge]{Leliaert_2018}
Jonathan Leliaert, Mykola Dvornik, Jeroen Mulkers, Jonas De~Clercq,
  MV~Milo{\v{s}}evi{\'c}, and Bartel Van~Waeyenberge.
\newblock Fast micromagnetic simulations on gpu—recent advances made with.
\newblock \emph{Journal of Physics D: Applied Physics}, 51\penalty0
  (12):\penalty0 123002, 2018.

\bibitem[Kumar and Adeyeye(2017)]{Kumar_2017}
D~Kumar and AO~Adeyeye.
\newblock Techniques in micromagnetic simulation and analysis.
\newblock \emph{Journal of Physics D: Applied Physics}, 50\penalty0
  (34):\penalty0 343001, 2017.

\bibitem[Abert et~al.(2013)Abert, Exl, Selke, Drews, and Schrefl]{Abert_2013}
Claas Abert, Lukas Exl, Gunnar Selke, Andr{\'e} Drews, and Thomas Schrefl.
\newblock Numerical methods for the stray-field calculation: A comparison of
  recently developed algorithms.
\newblock \emph{Journal of Magnetism and Magnetic Materials}, 326:\penalty0
  176--185, 2013.

\bibitem[Van~de Wiele et~al.(2008)Van~de Wiele, Olyslager, and
  Dupr{\'e}]{Van_de_Wiele_2008}
Ben Van~de Wiele, Femke Olyslager, and Luc Dupr{\'e}.
\newblock Application of the fast multipole method for the evaluation of
  magnetostatic fields in micromagnetic computations.
\newblock \emph{Journal of Computational Physics}, 227\penalty0 (23):\penalty0
  9913--9932, 2008.

\bibitem[Palmesi et~al.(2017)Palmesi, Exl, Bruckner, Abert, and
  Suess]{Palmesi_2017}
Pietro Palmesi, Lukas Exl, Florian Bruckner, Claas Abert, and Dieter Suess.
\newblock Highly parallel demagnetization field calculation using the fast
  multipole method on tetrahedral meshes with continuous sources.
\newblock \emph{Journal of Magnetism and Magnetic Materials}, 442:\penalty0
  409--416, 2017.

\bibitem[Vansteenkiste and Van~de Wiele(2011)]{Vansteenkiste_2011}
Arne Vansteenkiste and Ben Van~de Wiele.
\newblock Mumax: A new high-performance micromagnetic simulation tool.
\newblock \emph{Journal of Magnetism and Magnetic Materials}, 323\penalty0
  (21):\penalty0 2585--2591, 2011.

\bibitem[Ferrero and Manzin(2020)]{Ferrero_2020}
Riccardo Ferrero and Alessandra Manzin.
\newblock Adaptive geometric integration applied to a 3d micromagnetic solver.
\newblock \emph{Journal of Magnetism and Magnetic Materials}, page 167409,
  2020.

\bibitem[Livshitz et~al.(2009)Livshitz, Boag, Bertram, and
  Lomakin]{Livshitz_2009}
Boris Livshitz, Amir Boag, H~Neal Bertram, and Vitaliy Lomakin.
\newblock Nonuniform grid algorithm for fast calculation of magnetostatic
  interactions in micromagnetics.
\newblock \emph{Journal of Applied Physics}, 105\penalty0 (7):\penalty0 07D541,
  2009.

\bibitem[Exl and Schrefl(2014)]{Exl_2014}
Lukas Exl and Thomas Schrefl.
\newblock Non-uniform fft for the finite element computation of the
  micromagnetic scalar potential.
\newblock \emph{Journal of Computational Physics}, 270:\penalty0 490--505,
  2014.

\bibitem[Lepadatu(2019)]{Lepadatu_2019}
Serban Lepadatu.
\newblock Efficient computation of demagnetizing fields for magnetic
  multilayers using multilayered convolution.
\newblock \emph{Journal of Applied Physics}, 126\penalty0 (10):\penalty0
  103903, 2019.

\bibitem[Exl et~al.(2012)Exl, Auzinger, Bance, Gusenbauer, Reichel, and
  Schrefl]{Exl_2012}
Lukas Exl, Winfried Auzinger, Simon Bance, Markus Gusenbauer, Franz Reichel,
  and Thomas Schrefl.
\newblock Fast stray field computation on tensor grids.
\newblock \emph{Journal of computational physics}, 231\penalty0 (7):\penalty0
  2840--2850, 2012.

\bibitem[J.~García-Cervera(2014)]{j2014a}
Carlos J.~García-Cervera.
\newblock Numerical micromagnetics: A review.
\newblock \emph{Boletin de la Sociedad Espanola de Matematica Aplicada},
  39:\penalty0 130--135, 2014.

\bibitem[Rotarescu et~al.(2019)Rotarescu, Chiriac, Lupu, and
  Óvári]{rotarescu2019a}
C.~Rotarescu, H.~Chiriac, N.~Lupu, and T.~A. Óvári.
\newblock Micromagnetic analysis of magnetization reversal in fe77.5si7.5b15
  amorphous glass-coated nanowires.
\newblock \emph{Aip Advances}, 9\penalty0 (10):\penalty0 105316, 2019.
\newblock ISSN 21583226.
\newblock \doi{10.1063/1.5119450}.

\bibitem[Perez et~al.(2014)Perez, Torres, and Martinez-Vecino]{perez2014a}
Noel Perez, Luis Torres, and Eduardo Martinez-Vecino.
\newblock Micromagnetic modeling of dzyaloshinskii–moriya interaction in spin
  hall effect switching.
\newblock \emph{Ieee Transactions on Magnetics}, 50\penalty0 (11):\penalty0
  1--4, 2014.
\newblock ISSN 19410069, 00189464.
\newblock \doi{10.1109/TMAG.2014.2323707}.

\bibitem[Ragusa et~al.(2009)Ragusa, D'Aquino, Serpico, Xie, Repetto, Bertotti,
  and Ansalone]{ragusa2009a}
C.~Ragusa, M.~D'Aquino, C.~Serpico, B.~Xie, M.~Repetto, G.~Bertotti, and
  D.~Ansalone.
\newblock Full micromagnetic numerical simulations of thermal fluctuations.
\newblock \emph{Ieee Transactions on Magnetics}, 45\penalty0 (10):\penalty0
  5257303, 3919--3922, 2009.
\newblock ISSN 19410069, 00189464.
\newblock \doi{10.1109/TMAG.2009.2021856}.

\bibitem[Smith et~al.(2010)Smith, Nielsen, Christensen, Bahl, Bj{\o}rk, and
  Hattel]{Smith_2010}
Anders Smith, Kaspar~Kirstein Nielsen, DV~Christensen, Christian
  Robert~Haffenden Bahl, Rasmus Bj{\o}rk, and J~Hattel.
\newblock The demagnetizing field of a nonuniform rectangular prism.
\newblock \emph{Journal of Applied Physics}, 107\penalty0 (10):\penalty0
  103910, 2010.

\bibitem[Nielsen et~al.(2019)Nielsen, Insinga, and Bjørk]{nielsen2019a}
Kaspar~Kirstein Nielsen, Andrea~Roberto Insinga, and Rasmus Bjørk.
\newblock The stray- and demagnetizing field from a homogeneously magnetized
  tetrahedron.
\newblock \emph{I E E E Magnetics Letters}, 10:\penalty0 8918242, 2019.
\newblock ISSN 19493088, 1949307x.
\newblock \doi{10.1109/LMAG.2019.2956895}.

\bibitem[Sozer et~al.(2014)Sozer, Brehm, and Kiris]{sozer2014a}
Emre Sozer, Christoph Brehm, and Cetin~C. Kiris.
\newblock Gradient calculation methods on arbitrary polyhedral unstructured
  meshes for cell-centered cfd solvers.
\newblock \emph{52nd Aiaa Aerospace Sciences Meeting - Aiaa Science and
  Technology Forum and Exposition, Scitech 2014}, pages 24 pp., 24 pp., 2014.
\newblock \doi{10.2514/6.2014-1440}.

\bibitem[Insinga et~al.(2020)Insinga, Blaabjerg~Poulsen, Nielsen, and
  Bjørk]{insinga2020a}
A.R. Insinga, E.~Blaabjerg~Poulsen, K.K. Nielsen, and R.~Bjørk.
\newblock A direct method to solve quasistatic micromagnetic problems.
\newblock \emph{Journal of Magnetism and Magnetic Materials}, 510:\penalty0
  166900, 2020.
\newblock ISSN 18734766, 03048853.
\newblock \doi{10.1016/j.jmmm.2020.166900}.

\bibitem[Lopez-Diaz et~al.(2012)Lopez-Diaz, Aurelio, Torres, Martinez,
  Hernandez-Lopez, Gomez, Alejos, Carpentieri, Finocchio, and
  Consolo]{Lopez_Diaz_2012}
L~Lopez-Diaz, D~Aurelio, L~Torres, E~Martinez, MA~Hernandez-Lopez, J~Gomez,
  O~Alejos, M~Carpentieri, G~Finocchio, and G~Consolo.
\newblock Micromagnetic simulations using graphics processing units.
\newblock \emph{Journal of Physics D: Applied Physics}, 45\penalty0
  (32):\penalty0 323001, 2012.

\bibitem[Leliaert and Mulkers(2019)]{Leliaert_2019}
Jonathan Leliaert and Jeroen Mulkers.
\newblock Tomorrow’s micromagnetic simulations.
\newblock \emph{Journal of Applied Physics}, 125\penalty0 (18):\penalty0
  180901, 2019.

\bibitem[Bjørk and Nielsen(2019)]{MagTense}
R~Bjørk and K.~K. Nielsen.
\newblock Magtense - a micromagnetism and magnetostatic framework.
\newblock \emph{doi.org/10.11581/DTU:00000071, https://www.magtense.org}, 2019.

\bibitem[Kr{\"u}ger et~al.(2013)Kr{\"u}ger, Selke, Drews, and
  Pfannkuche]{Kruger_2013}
Benjamin Kr{\"u}ger, Gunnar Selke, Andr{\'e} Drews, and Daniela Pfannkuche.
\newblock Fast and accurate calculation of the demagnetization tensor for
  systems with periodic boundary conditions.
\newblock \emph{IEEE transactions on magnetics}, 49\penalty0 (8):\penalty0
  4749--4755, 2013.

\bibitem[Lebecki et~al.(2008)Lebecki, Donahue, and Gutowski]{Lebecki_2008}
Krzysztof~M Lebecki, Michael~J Donahue, and Marek~W Gutowski.
\newblock Periodic boundary conditions for demagnetization interactions in
  micromagnetic simulations.
\newblock \emph{Journal of Physics D: Applied Physics}, 41\penalty0
  (17):\penalty0 175005, 2008.

\bibitem[Moskowitz and Della~Torre(1966)]{Moskowitz_1966}
R~Moskowitz and E~Della~Torre.
\newblock Theoretical aspects of demagnetization tensors.
\newblock \emph{IEEE Transactions on Magnetics}, 2\penalty0 (4):\penalty0
  739--744, 1966.

\bibitem[NIS()]{NIST_2020}
$\mu$mag standard problems, accessed on november 1st, 2020.
\newblock URL \url{https://www.ctcms.nist.gov/~rdm/toc.html}.

\bibitem[Streibl et~al.(1999)Streibl, Schrefl, and Fidler]{Streibl_1999}
B~Streibl, T~Schrefl, and J~Fidler.
\newblock Dynamic fe simulation of $\mu$mag standard problem no. 2.
\newblock \emph{Journal of applied physics}, 85\penalty0 (8):\penalty0
  5819--5821, 1999.

\bibitem[Donahue et~al.(2000)Donahue, Porter, McMichael, and
  Eicke]{Donahue_2000}
Michael~J Donahue, Donald~G Porter, Robert~D McMichael, and J~Eicke.
\newblock Behavior of $\mu$mag standard problem no. 2 in the small particle
  limit.
\newblock \emph{Journal of Applied Physics}, 87\penalty0 (9):\penalty0
  5520--5522, 2000.

\bibitem[Shepherd et~al.(2014)Shepherd, Miles, Heil, and
  Mihajlovi{\'c}]{Shepherd_2014}
David Shepherd, Jim Miles, Matthias Heil, and Milan Mihajlovi{\'c}.
\newblock Discretization-induced stiffness in micromagnetic simulations.
\newblock \emph{IEEE Transactions on Magnetics}, 50\penalty0 (11):\penalty0
  1--4, 2014.

\bibitem[Ramstock et~al.(1996)Ramstock, Hubert, and Berkov]{Ramstock_1996}
Klaus Ramstock, Alex Hubert, and D~Berkov.
\newblock Techniques for the computation of embedded micromagnetic structures.
\newblock \emph{IEEE Transactions on Magnetics}, 32\penalty0 (5):\penalty0
  4228--4230, 1996.

\bibitem[Hirano and Hayashi(1999)]{Hirano_1999}
Satoshi Hirano and Nobuo Hayashi.
\newblock Multigrid computation for micromagnetics.
\newblock \emph{Journal of applied physics}, 85\penalty0 (8):\penalty0
  6205--6207, 1999.

\bibitem[Garcia-Cervera and Roma(2006)]{Garcia_2006}
Carlos~J Garcia-Cervera and Alexandre~M Roma.
\newblock Adaptive mesh refinement for micromagnetics simulations.
\newblock \emph{IEEE transactions on magnetics}, 42\penalty0 (6):\penalty0
  1648--1654, 2006.

\bibitem[Sun and Monk(2006)]{Sun_2006}
Jiguang Sun and Peter Monk.
\newblock An adaptive algebraic multigrid algorithm for micromagnetism.
\newblock \emph{IEEE transactions on magnetics}, 42\penalty0 (6):\penalty0
  1643--1647, 2006.

\end{thebibliography}
\end{document}